\documentclass[sigplan,nonacm]{acmart} %
\acmSubmissionID{137}
\renewcommand\footnotetextcopyrightpermission[1]{}
\usepackage[utf8]{inputenc}
\usepackage{array}
\newcolumntype{?}{!{\vrule width 2pt}}
\usepackage{subcaption}
\usepackage{xcolor}
\usepackage[binary-units]{siunitx}
\usepackage{booktabs}
\usepackage{makecell}
\usepackage{multirow}
\usepackage{calc}
\usepackage{xspace}
\usepackage{tikz}
\usepackage{svg}
\usetikzlibrary{calc}
\usetikzlibrary{fit}
\usetikzlibrary{positioning}
\usetikzlibrary{shapes.symbols}

\copyrightyear{V}

\DeclareSIUnit{\nothing}{\relax}
\DeclareSIUnit{\x}{\times}

\newcommand{\Boxer}{{Boxer}\xspace}

\newcommand{\sref}[1]{\S\ref{#1}}

\begin{document}
\date{}
\title[Boxer: FaaSt Ephemeral Elasticity for Off-the-Shelf Cloud Applications]{Boxer: FaaSt Ephemeral Elasticity for \\ Off-the-Shelf Cloud Applications
}
\author{
Michael Wawrzoniak$^1$,
Rodrigo Bruno$^2$,
Ana Klimovic$^1$,
Gustavo Alonso$^1$ \\
$^1$Systems Group, Dept. of Computer Science, ETH Zurich  \\
$^2$INESC-ID/Técnico, U. Lisboa
}

\maketitle
\pagestyle{plain}

\section*{Abstract}
Elasticity is a key property of cloud computing. However, elasticity is offered today at the granularity 
 of virtual machines, which take tens of seconds to start. This is insufficient to react to load spikes and sudden failures in latency sensitive applications, leading users to resort to expensive overprovisioning. Functions as a Service (FaaS) %
provides significantly higher elasticity than VMs, but %
comes coupled 
with an event-triggered programming model and a constrained execution environment that makes them
unsuitable for off-the-shelf %
applications. Previous work %
tries to 
overcome these obstacles but often 
requires re-architecting the applications. In this paper, we show how off-the-shelf applications can transparently benefit from \textit{ephemeral elasticity} with FaaS. %
We built \Boxer, an interposition layer spanning VMs and AWS Lambda, that intercepts application execution and emulates the network-of-hosts environment that applications expect when deployed in a conventional VM/container environment.
The ephemeral elasticity of \Boxer enables significant performance and cost savings for off-the-shelf applications with, e.g., recovery times over 5x faster than EC2 instances 
and absorbing load spikes comparable to overprovisioned EC2 instances.

\section{Introduction}

\begin{figure}[t]
  \centering
  \includegraphics[width=1.0\linewidth]{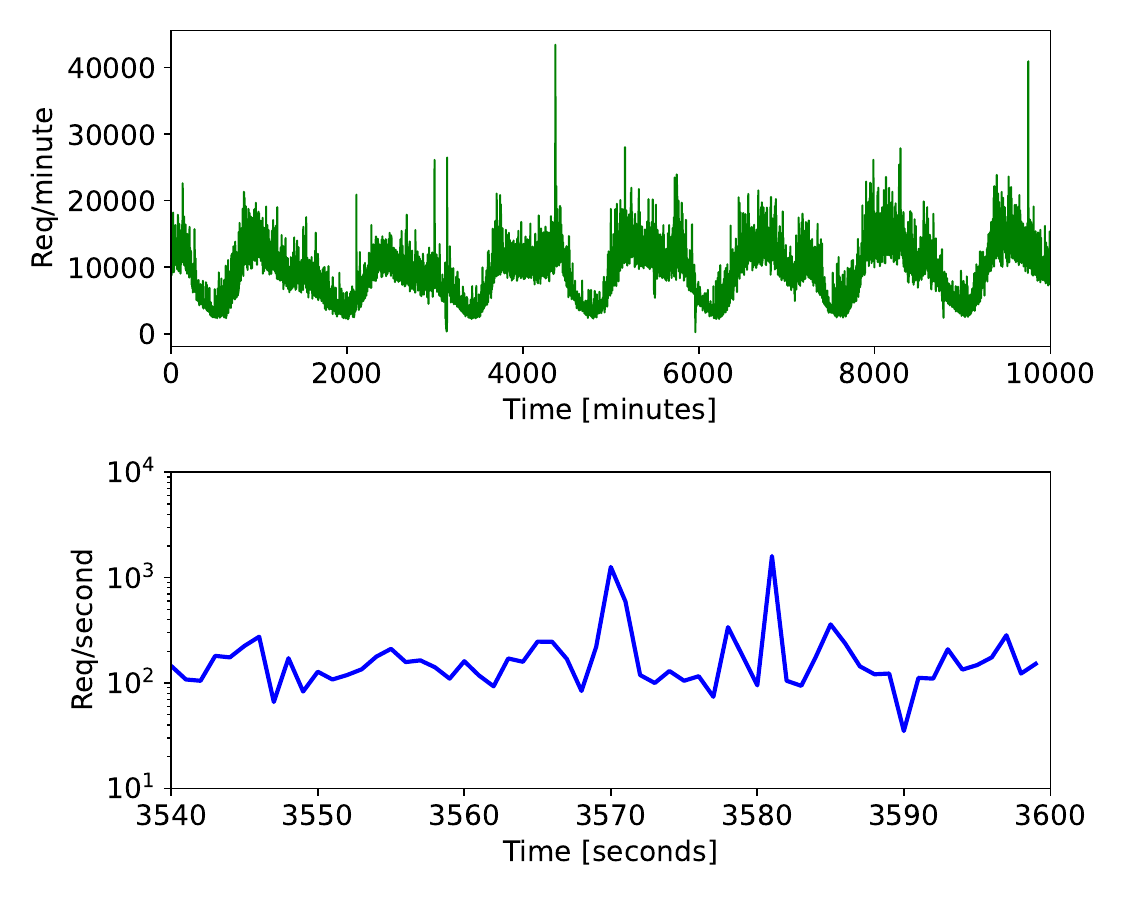}
  \caption{Reddit requests over 7 days (top) and 1 minute (bottom). Extracted from a public Reddit 2015 trace.}
  \label{fig:eval_reddit}
\end{figure}

\begin{figure*}[t]
  \centering
  \begin{minipage}[t]{.495\textwidth}
    \includegraphics[keepaspectratio,width=\textwidth]{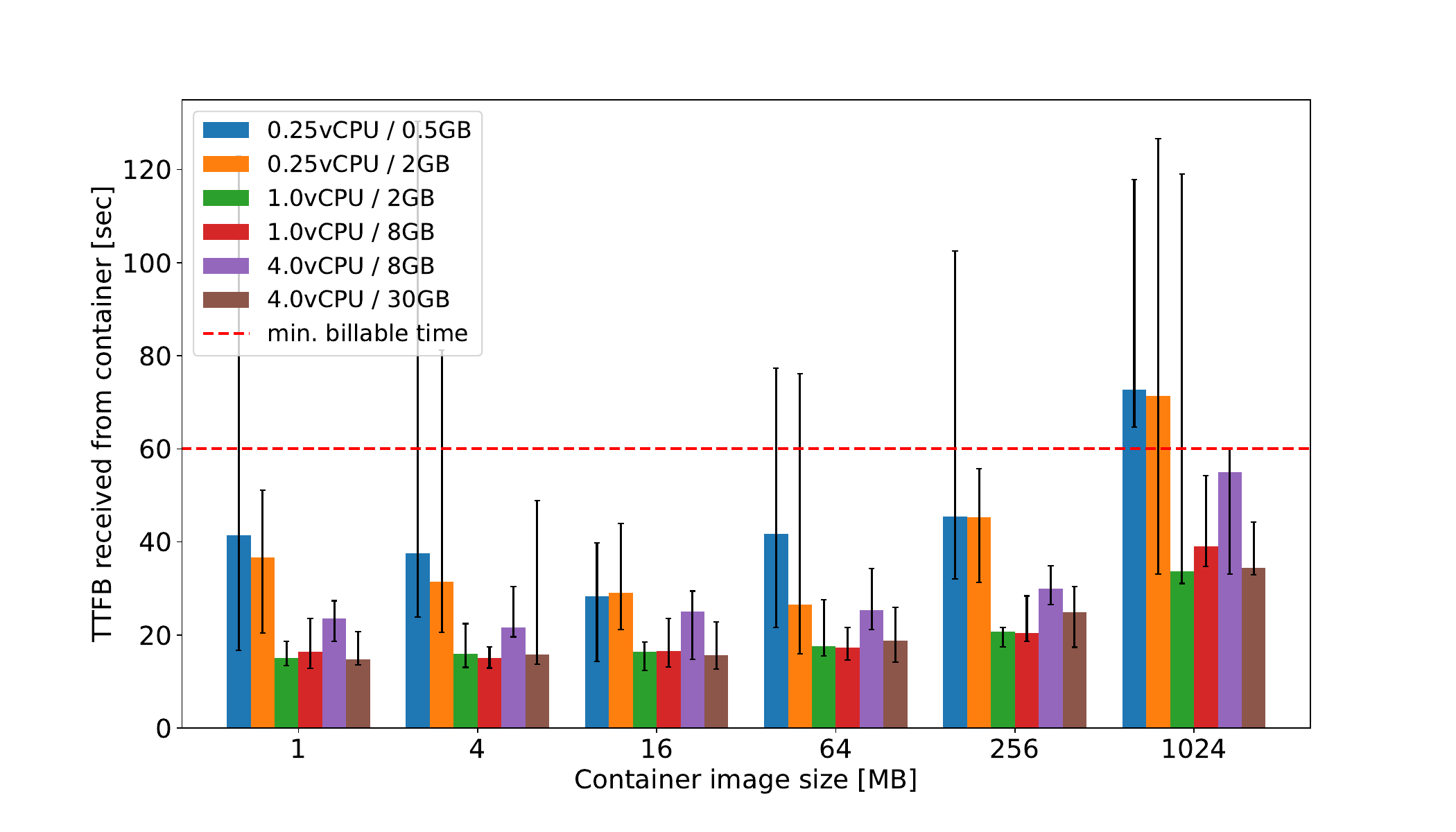}
    \centering
    (a) AWS Fargate/ECS container service
  \end{minipage}                     
  \begin{minipage}[t]{.495\textwidth}
    \includegraphics[keepaspectratio,width=\textwidth]{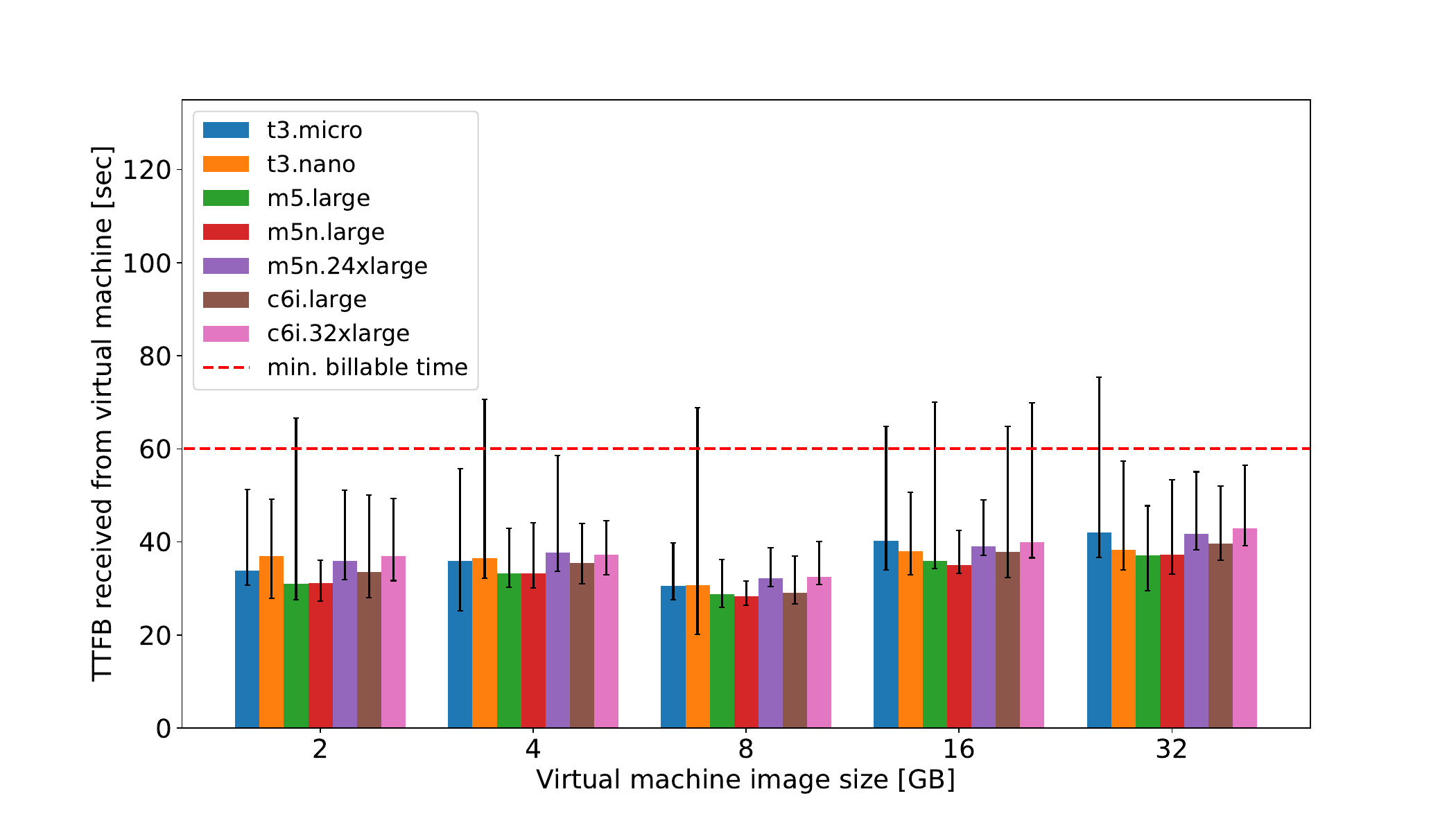}
    \centering
    (b)  AWS EC2 virtual machines
  \end{minipage}
  \caption{Median instantiation times of containers and VMs services, error bars are min and max values. Details in Section~\ref{sec:vmc-elasticity}}
  \label{fig:motivation_vm_startup}
\end{figure*}

Elastic resource allocation is a key feature of cloud computing~\cite{View-on-Cloud}. Cloud users rent virtual machines (VMs) on-demand to meet the resource requirements of their applications. 
However, the elasticity granularity offered by today's virtual machines is insufficient to react to sudden load spikes or VM failures that latency-sensitive cloud applications commonly experience. For example, Figure~\ref{fig:eval_reddit} shows that the request rate for a Reddit web service application varies up to \textit{two orders of magnitude} within \textit{five seconds}. Meanwhile, instantiating just a VM or allocating new resources for a 'fast starting' container in the cloud takes \textit{tens of seconds} (Figure~\ref{fig:motivation_vm_startup}).

Since conventional cloud infrastructure is slow to respond when load spikes or failures occur, users often resort to overprovisioning resources to provide the illusion of higher elasticity~\cite{borg-eurosys20, quasar, resource-central-sosp17, mlaas-nsdi22}. This is expensive for users as they rent and pay for more and/or larger VMs than they really need. Widespread overprovisioning is also costly for cloud providers, who despite techniques like over-committing and harvesting slack resources~\cite{borg-eurosys20, overcommitment, harvestVM}, still need to power significantly more machines than necessary to support the aggregate load~\cite{alibaba-study}. For example, the 2019 Borg traces show that CPU and memory utilization is only $\sim$60\%, even when the provider overcommits resources~\cite{borg-eurosys20}.

Function as a Service (FaaS) platforms, such as Azure Functions~\cite{azurefunctions} and AWS Lambda~\cite{awslambda}
offer highly elastic compute pools that %
automatically scale based on the number of tasks that users invoke. 
The ``serverless'' execution model of these platforms simplifies resource allocation, scales resources on demand, and offers fine-grained billing favorable for short tasks. Under the hood, FaaS platforms execute tasks in lightweight VMs designed to boot quickly (e.g., 100s of milliseconds~\cite{Firecracker}). %
However, although it offers high elasticity, current FaaS cloud platforms couple the %
fast-booting VMs with an event-triggered programming model and a constrained execution environment that makes them 
unfit to run general-purpose cloud applications off-the-shelf~\cite{HellersteinCIDR19}. 
Existing FaaS platforms, force applications to be written as collections of short-lived, stateless functions, which cannot accept network connections while executing~\cite{HellersteinCIDR19,RiseofServerless19,Berkeley-CACM,kappa}.

Previous work %
tries to overcome these obstacles by implementing point solutions for various use cases, such as data analytics~\cite{Lambada, starling}, stream processing~\cite{sponge}, video processing~\cite{ExCamera}, and machine learning training~\cite{ServerlessML21}.
Other solutions address some of the limitations of FaaS (e.g., function-to-function communication) but still require %
re-architecting large software stacks to adapt to the FaaS programming model~\cite{unum, anna, pocket, gg, Boxer-CIDR21, fmi}.

In this work, we show how off-the-shelf cloud applications can transparently benefit from \textit{ephemeral elasticity} with FaaS. We focus on ephemeral usage of FaaS to absorb load spikes and accommodate sudden failure recovery, rather than running an entire application from start to end since traditional long-running VMs still provide a cost advantage compared to FaaS~\cite{prime-video-not-serverless,Lambada,starling} and are suitable to serve steady application load. Our aim is to seamlessly run applications across traditional long-running VMs and FaaS instances without requiring changes to applications.  
The key challenge is providing a familiar distributed programming model (i.e., POSIX-style network-of-hosts) --- which generic cloud applications expect --- on top of existing FaaS platforms. We achieve this by designing an interposition layer (now available to users, eventually supported by the cloud provider) deployed on top of existing FaaS platforms to emulate the necessary network, file system, and name resolution functionality.  %
We have implemented such interposition layer, \textit{\Boxer}, on top of AWS Lambda. \Boxer does not require changes to the application and integrates with traditional infrastructure orchestration tools such as Docker Compose. 
\Boxer intercepts system C Library function calls and emulates the necessary network-of-hosts environment (network, file system, name resolution) that applications expect when deployed in a VM/container environment. We show that \Boxer can be used to quickly absorb load bursts in an unmodified microservice application (DeathStar benchmark).%
Similarly, we show how an unmodified Zookeeper quorum running on EC2 can be quickly restored by replacing a failed node with a Lambda instance in about 6 seconds while doing the same with VMs takes close to one minute. These results demonstrate the potential of \Boxer to provide ephemeral elasticity in a transparent manner.

\section{The elasticity dream: are we there yet?}

To characterize the elasticity requirements of cloud applications and understand the limitations of VM and container-based cloud infrastructure, we analyze a public Reddit trace~\cite{redditdataset} that includes user requests per second %
as an example of a web-based microservice application. From the dataset we extract two subsets: a 7-day trace containing the number of requests per minute, and a 1-hour trace containing the number of requests per second (Figure~\ref{fig:eval_reddit}), from which we draw two key observations:

\textbf{Observation \#1:} the 7-day trace (bottom plot in Figure~\ref{fig:eval_reddit}) displays an evident daily pattern for which the infrastructure can be scaled over the course of minutes and hours. For such course-grained load variations, high infrastructure elasticity is not required;

\textbf{Observation \#2:} when looking at the 1-minute trace, we find significant workload burstiness. Unlike the 7-day trace, the 1-minute trace requires highly elastic or highly overprovisioned infrastructure to be able to serve workload changes of more than an order of magnitude in a few seconds. 

We conclude that real workloads benefit from two different elasticity granularities: coarse-grain elasticity to scale the entire infrastructure over the period of minutes and hours, and fine-grain elasticity to serve unpredictable user request bursts at the second scale. Next, we demonstrate that no existing cloud infrastructure can cost-efficiently satisfy both types of elasticity. To bridge this gap, we propose the idea of \textit{ephemeral elasticity}, a solution for affordable and highly elastic cloud infrastructure.

\subsection{Virtual Machine and Container Elasticity}
\label{sec:vmc-elasticity}

Conventional virtual machine or container service deployments offered by cloud providers have significant, often deeply intertwined, inefficiencies. In 
Figure~\ref{fig:motivation_vm_startup} we explore the issue through experiments that measure the time to first byte (TTFB) received for different image sizes, container resources sizes (vCPU, memory) and virtual machine types. We measure the time from issuing a local (in the same availability zone and VPS) instantiation request to receiving back the first one-byte UDP packet sent from the instantiated purpose-built minimal container/virtual machine image. The experiment is repeated 10 and 32 times for each of the ECS and EC2 configurations, respectively. As the data shows, real-world VM and container services, such as AWS EC2 and AWS Fargate, take on the order of 10s of seconds to allocate new resources, initialize, and return the first byte of data to a user. And that without including the additional time needed to instantiate the application intended to run on the VM.
Note that although containers can be 'fast starting,' the cloud services providing them (AWS Fargate) still need to allocate additional resources for them, adding to the container instantiation times.
This long initialization time 
makes it difficult for %
applications to respond quickly to unpredictable load spikes or node failures. As a consequence, users commonly %
overprovision resources, which underutilizes expensive hardware infrastructure, e.g., memory utilization in cloud deployments is typically between 50 and 55\%~\cite{googletraceanalysis, rose} and very rarely exceeds 80\%~\cite{imbalance}. 

From here we conclude that VM and container-based deployments are suitable for slowly evolving loads (that changes in minutes and hours, e.g., in the 7-day trace in Figure~\ref{fig:eval_reddit}), but not for high, unpredictable load bursts (e.g., in the second range seen in the 1-minute trace of Figure~\ref{fig:eval_reddit}).

\subsection{Ephemeral Elasticity}
\label{sec:ephemeral}

We propose the concept of \textit{ephemeral elasticity}:  running applications across both VM/containers as well as FaaS. The goal is to have a single orchestrated application deployment that takes advantage of both types of infrastructure: one for predictable load, and the other for request bursts.  In this section, we estimate the potential of ephemeral elasticity to reduce resource overprovisioning and reduce the cost of running applications on the cloud. To do so, we conduct a cost analysis to compare using AWS Lambda to accommodate load bursts with overprovisioned AWS EC2 VMs. %
The cost of each deployment, including the cost of the EC2 baseline infrastructure and Lambdas for accommodating bursts, can be calculated as follows:
\[ \sum_{t=0}^{T} \Biggr[ \frac{\beta}{\alpha} \times \$\textsubscript{EC2}  + max\Big(0, \frac{\delta\textsubscript{t} - \beta}{\gamma} \times \$\textsubscript{Lambda}\Big) \Biggr] \]

where $\beta$ is the number of requests served by EC2 VMs; $\alpha$ and $\gamma$ the throughput per core of EC2 and Lambda (measured for Deathstar microservice in \sref{sec:eval:deathstar}). 
\$\textsubscript{EC2} and \$\textsubscript{Lambda} are the cost per second per core 
(we base on \texttt{c6g.2xlarge} VM and a 2GB Lambda);
$\delta\textsubscript{t}$ the load (number of requests) at time $t$.

\begin{figure}[t]
  \centering
  \includegraphics[width=1.0\linewidth]{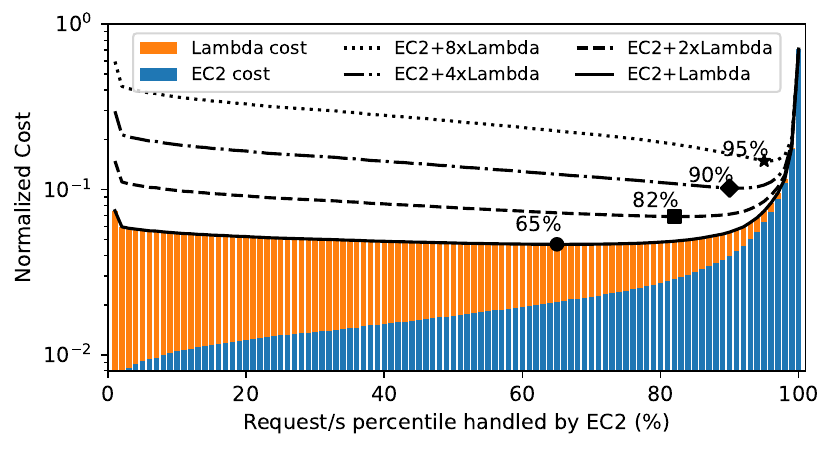}
  \includegraphics[width=1.0\linewidth]{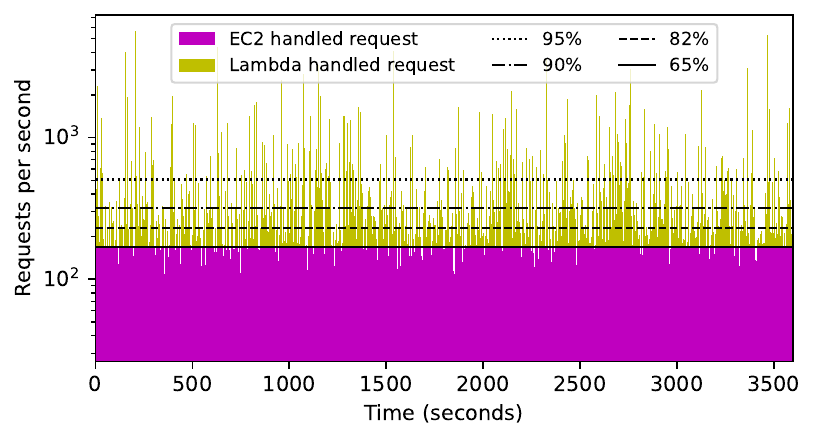}
  \caption{Reddit deployment cost (top) for different EC2 capacities using Lambda to handle requests that exceed capacity. 1-day Reddit trace (bottom) showing requests handled by EC2 and Lambda to minimize cost while providing capacity to handle all requests (c100), 65\% of total requests corresponds to the level of 3\% of the observed maximum. (Section~\ref{sec:ephemeral})}
  \label{fig:eval_reddit2}
\end{figure}

\begin{table}
    \centering
    \begin{tabular}{r|cccc}
         & c100 & c99 & c95  & c90 \\
         \hline 
       EC2 + Lambda    & 93.42\% & 75.53\% & 43.40\%   & 21.86\%\\
       EC2 + 2xLambda  & 90.31\% & 65.03\% & 25.71\%   & 5.87\%\\
       EC2 + 4xLambda  & 85.60\% & 50.08\% & 7.17\%    & no-saving\\
       EC2 + 8xLambda  & 78.95\% & 31.35\% & no-saving & no-saving\\
    \end{tabular}
    \caption{Estimated cost savings relative to different EC2 provisioning levels (c100, c99, c95, c90) based on Reddit trace.} 
    \label{tab:saving}
\end{table}

Figure~\ref{fig:eval_reddit2} (top) presents the normalized total deployment cost per hour for the Reddit trace with a varying percentage of capacity served by EC2 instances ($\beta$ goes from 0 to the maximum number of requests at any moment). If no capacity is handled by EC2, then all requests are served by Lambda instances, leading to a high cost per hour. On the other hand, if all requests are handled by EC2, a significant amount of overprovisioning is required to handle all request bursts, leading to a high cost. 
The deployment that minimizes cost and therefore resource overprovisioning is obtained by combining EC2 and Lambda instances (approximately 65\% of the capacity handled by EC2), thereby demonstrating the potential of the ephemeral elasticity idea. If more Lambda resources are necessary to process requests (because of inflexible resource allocation options, additional memory, or networking requirements,) total cost increases and best capacity allocation shifts (e.g., 82\% for 2x Lambda per-request requirements.) Figure~\ref{fig:eval_reddit2} (bottom) ilustrated the best capacity allocation between EC2 and Lambda at 65\% level, when the request rate is below 65\% of the maximum the long-running VM capacity handles it, when it is above, additional ephemeral capacity based on Lambda is used to scale up temporairly. Table~\ref{tab:saving} shows estimated cost reduction when using ephemeral elasticity relative to different levels of EC2 VM overprovisioning; even when EC2 is provisioned to handle only 95\% of maximum requests per second (c95), and 2 $\times$ Lambdas are needed to service requests, the estimated cost reduction is over 25\%.

\subsection{Incompatible Datacenter Execution Models}

Ephemeral elasticity is not available today as FaaS, and VMs and containers, operate under two different models:

\textbf{Network-of-hosts} is the classic and dominant datacenter model that has been in use for decades. Systems are built around the concept of long-running interconnected hosts that execute a collection of application processes that communicate via the network. Processes can be reached using various addresses and names, predominantly based on IP addresses, hostnames, and L4 port numbers.

\textbf{Event-triggered functions} is the model promoted by FaaS services. Systems are composed of a collection of functions executed in response to events. These functions can be composed into complex systems by arranging them into event-driven dataflow graphs. With their fast starting time and high invocation parallelism, they are suitable to highly burstable loads as the one depicted in Figure~\ref{fig:eval_reddit}.
However, functions are expected to be stateless, have limited networking, and limited execution time (e.g., up to 15 minutes in Amazon Lambda and 30 minutes in Azure Functions)~\cite{HellersteinCIDR19,Berkeley-CACM,Boxer-CIDR21} and are more expensive than a similar VM instances.

\textit{Takeaway:} The mismatch between execution models prevents existing cloud applications from transparently and efficiently combining FaaS and VM/container infrastructure to achieve ephemeral elasticity.

\section{System Requirements and Assumptions}
\label{sec:requirerments}
To achieve the vision of fast ephemeral elasticity, we identify the following requirements for our \Boxer prototype: %

\textbf{Work with existing platforms:}
To test feasibility and to avoid making unrealistic assumptions, \Boxer should run on a cloud platform available today. It must not assume any more privileges than those available to a regular tenant of a publicly available cloud platform. For example, we use AWS EC2 as the virtual machine platform and AWS Lambda FaaS as the serverless platform.

\textbf{Application transparency:}
 To make the technique general and broadly applicable to a large set of existing applications, \Boxer must not rely on modifying or specializing applications.
When \Boxer runs generic applications in environments for which they were not designed (e.g., FaaS), the environment must be transparently emulated to match what unmodified applications expect. 

\textbf{Efficiency:}
Any introduced overhead (e.g., to emulate some aspects of the execution environment) must be sufficiently small to not violate the performance objectives or timing constraints of the user application. 
For example, the system must provide network connections to the application faster than its connection timeout, to avoid getting trapped in connect-retry loops.
Beyond not violating such assumptions, reducing system overhead is particularly important as FaaS instances can be very small; this makes system overhead play a proportionally larger role. %

\textbf{Orchestration compatibility:}
Cloud application deployments depend on orchestration systems. 
If \Boxer required a new or modified orchestration system, it would reduce its usability and generality. 
Thus, there should be a way to use \Boxer with unmodified popular orchestration systems, such as Docker Compose~\cite{docker-compose}, to pave a path to adoption in practice.

\smallskip

We assume that \Boxer and the application are used by the same tenant so that there is no incentive for the application code to escape the mechanism used by \Boxer. We assume that the guest applications are \emph{cooperative}, we do not aim to prevent applications from circumventing the mechanisms we provide
Second, we assume that individual sub-services of the target applications (e.g., individual worker nodes, quorum members nodes, microservice nodes) can be scaled up and down by adding and removing service nodes. This is a standard paradigm in microservice architectures.
Third, we assume that the application's long-term persistent state is either stored in the long-running VMs or in remote cloud storage, not in the short-lived ephemeral workers. %
Lastly, our current system prototype does not aim to provide complete transparency (e.g., applications can find out they are running in \Boxer) however, we do not aim to cover unusual corner cases for now as the functionality of typical applications is not affected.

\begin{figure}[t]
  \centering
  \includegraphics[width=1.07\linewidth]{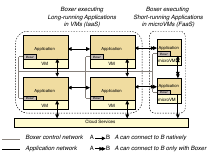}
  \caption{An unmodified long-running networked datacenter application running in VMs (temporarily) augmented with FaaS microVM based ephemeral elasticity using \Boxer.}
  \label{fig:ephemeral-elasticity}
\end{figure}

\section{\Boxer System Design}
\label{sec:Boxer}

We provide an overview of the \Boxer system and the reasoning that guided our design of such an interposition layer.

\subsection{Design Overview}

\Boxer must emulate the required \emph{network-of-hosts} execution model to applications on top of the \emph{event-triggered-functions} model of the serverless platform.
Since neither applications nor platforms can be modified, %
we design \Boxer as an interposition layer between cloud applications and platforms, i.e., a form of \textit{cloud overlay} (Figure~\ref{fig:ephemeral-elasticity}).
To achieve this, \Boxer \emph{intercepts} the execution of the guest applications running on top of it, and uses the platform resources below to \emph{emulate} the expected environment for the application.

\subsection{\bf Intercepting the guest application execution}
\label{sec:execution}
Interposition must be enforced dynamically at runtime and not by modifying the application code or binaries.
At the same time, our target platforms include microVMs of publicly available FaaS services (AWS Lambda) that provide a restricted environment.
These restrictions eliminate approaches to trap application executions that rely on hardware-accelerated nested virtualization, modifying the kernel, or loading kernel modules.
On the other hand, unprivileged userspace virtualization techniques based on full dynamic translation (e.g., QEMU) add too much overhead to be viable. %
We also cannot rely on other standard methods to intercept system calls of Linux processes, such as those based on \texttt{ptrace} or \texttt{seccomp} system calls, 
because their use is restricted in the unprivileged environment. %

Given this combination of constraints, we choose to intercept the application execution at the system C Library function call level.
We implement the interception of the calls by controlling the dynamic linking of application processes as they started (described in~\sref{sec:boxer:supervisor}).
Hence, our current system targets applications that dynamically link with the system C Library and do not directly issue system calls that \Boxer must intercept.

Compared to the other trapping approaches, interposition at the function call level incurs a negligible performance penalty of an additional function call, supporting our requirement for low system overhead. 
To minimize the overall performance overhead of the emulation, \Boxer aims to limit the intercepted surface area to a minimum.
We designed the system to reduce the number of intercepted functions and to delegate as much functionality directly to the underlying platform as possible.
This also improves the fidelity of the emulation since fewer mechanisms must be re-implemented.
In particular, \Boxer leaves signals and memory management directly to the underlying platform, and the interception of system C Library calls are limited to the control path operations only.
Most significantly, we avoided intercepting data path calls (e.g. \texttt{send}, \texttt{write}, \texttt{recv}, \texttt{read}, \texttt{sendfile}) and I/O notification calls (e.g. \texttt{epoll}, \texttt{select}) providing no-overhead performance for those operations.

\begin{figure}[t]
  \centering
  \includegraphics[width=1.0\linewidth]{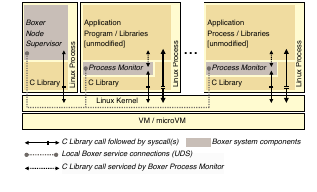}
  \caption{\Boxer node (VM, microVM, or a container).} 
  \label{fig:boxer-node}
\end{figure}

Figure \ref{fig:boxer-node} shows the components running on each node in a distributed application cluster with \Boxer. The \Boxer Process Monitor (\sref{sec:boxer:processmonitor}) is the system component that is responsible for intercepting the necessary system C Library calls. 
It is loaded by the Node Supervisor (described further in \sref{sec:boxer:supervisor}) into every guest application process.
The Process Monitor is limited to a thin shim that interacts with the local Node Supervisor that provides services needed for the emulation.

\subsection{\bf Emulating the network-of-host model}

Emulating the \emph{network-of-hosts} model that spans all nodes participating in a \Boxer setup (VMs, containers, microVMs of FaaS) requires providing additional services to the guest applications.
\Boxer exposes all nodes, including FaaS microVMs as networked hosts to the guest application.
To provide network connectivity between different hosts, \Boxer must provide network transport between the hosts, manage network addresses, and provide hostname resolution, all these services are provided by the Node Supervisor.

This separation of functionality between stateless and thin Process Monitors and a Node Supervisor providing the services to all local processes requires a communication channel capable of request-response commands, sending file descriptors, and signaling some I/O notifications without interfering with the guest application.
For the former two, we chose to use Unix domain sockets because they can be used to send file descriptors between processes.
For the later, we developed a technique of delivering the required signals to Process Monitors using marked local stream sockets (we later refer to them as \emph{signal sockets}) that requires minimal mechanism in the Process Monitor and is compatible with the guest using blocking and non-blocking I/O.

\section{System Implementation}
In the following sections we discuss some implementation aspects of the Process Monitor (PM),
the Node Supervisor (NS),
and how they interact to provide the environment emulation, what services are provided and how the system can be transparently used with container orchestration tools~(\sref{sec:boxer:deployment}).

\subsubsection*{Process Monitor (PM)}
\label{sec:boxer:processmonitor}
\Boxer PM is responsible for selectively intercepting the execution of guest application processes to emulate the desired environment.
Every guest process as it is loaded to be executed by \Boxer is first dynamically linked with the PM library.
The library exports a set of symbols that are normally exported by the system C Library, and it is linked between the platform system C Library and the application program and application libraries (Figure ~\ref{fig:boxer-node}). 
This provides an interposition layer where the PM library can intercept the guest application execution by selectively intercepting system C Library calls made by the guest application processes.
Currently, the intercepted  calls are for stream socket (\texttt{socket}, \texttt{bind}, \texttt{connect}, \texttt{listen}, \texttt{accept}), 
network address and hostname resolution (\texttt{getaddrinfo}, \texttt{uname}), 
and file access (\texttt{open}, \texttt{close}).
Together with their variants and companion functions, a total of 24 functions are intercepted.
Notably, the selective interception avoids intercepting any data path functions (such as \texttt{read}, \texttt{recv}, \texttt{send}, \texttt{write}, etc.) or I/O notification functions (such as \texttt{epoll}, \texttt{select}, etc.).
The intervention in the execution of the guest processes is limited to control plane operations for establishing network connections and network and file system naming - 
there is no additional overhead once network connections are established or files are opened.

The functionality in the PM is kept to a minimum, with no state persisted between intercepted calls, and with most of the functionality accessed through the NS services. 
PMs access their local NS by exchanging messages on a named Unix domain socket, referred to as \emph{service connections}.
For some of the intercepted calls, e.g., \texttt{getaddrinfo}, the functionality of PM is limited to just parsing the arguments, sending an appropriate request message (in this case \texttt{NameLookupReq}) to the NS
and then formatting and returning the results to the caller.
For other intercepted functions, the procedure before sending a request on service connections is more involved.
For example, the protocol requires that when handing an intercepted \texttt{accept} function, a non-blocking native accept call must be made before potentially proceeding with sending an accept request to \Boxer.
This is because if \Boxer has a ready connection to be accepted by the guest program, and no process is blocked to accept it, the program may be waiting for the I/O event to be delivered before calling accept (e.g. via epoll).
To address such a scenario, \Boxer NS will create a \emph{signal connection}, a local connection from a reserved address only to trigger a matching I/O event delivery to the guest program.
If the guest process then calls accept, the PM will first accept the signal connection, discard it, and then proceed to send the service request to the NS to receive the new socket from \Boxer (as a file descriptor sent over the service connection) and then return it to the guest program as the return value from the original accept call. %
Handling \texttt{bind} and \texttt{connect} also requires substantial setup before issuing service requests, but the main mechanisms ara implemented as services, leaving the PM stateless and relatively simple.

\subsubsection*{Node Supervisor (NS)}
\label{sec:boxer:supervisor}

The NS is an unprivileged process that runs in every node (VM, container, or microVM) participating in the \Boxer network (Figure \ref{fig:boxer-node}).
The NS is responsible for managing the local guest application processes, servicing the requests of the local PMs, and maintaining the control network with NSs of other nodes in the network.

The NS starts the specified guest application programs (with the specified arguments and additional configuration environment variables) and preloads all of their processes with the PM library (\sref{sec:boxer:processmonitor}).
It then listens for the PMs to open local service connections and start issuing requests.

The secondary role of the NS is to bootstrap and maintain a control network with other remote NSs.
The control network is used to send and receive commands between remote nodes, including network setup requests. 
Currently, the control network is based on direct TCP connectivity that supervisors establish between nodes when they start.
Internally, the supervisor forwards the local and remote requests to one of its local services, such as the networking or coordination service discussed in the following sections.

\subsubsection*{Network Service}
\label{sec:boxer:net}
\begin{figure}[t]
   \centering
  \includegraphics[width=1.06\linewidth]{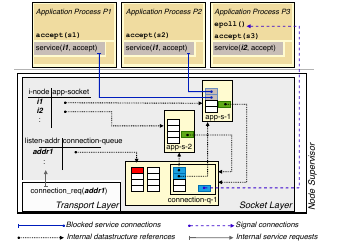}
  \caption{A configuration of core internal data structures of the stream socket layer for listening sockets (~\sref{sec:boxer:net}).}
  \label{fig:listen-socket}
\end{figure}

Conceptually, the network service is composed of two layers: the socket layer and the transport layer.
The socket layer provides the mechanism for creating and setting up guest application sockets.
The network transport layer (not meant to be interpreted with OSI model) provides network data delivery between \Boxer nodes that can back the sockets created by the socket layer.
\\

\noindent
\emph{Socket Layer}
The socket layer interacts with the PMs from above and with the transport layer from below.
When a PM intercepts a relevant socket call, e.g., \texttt{connect} call to establish a stream connection to a destination in \Boxer network, it will send a request to the network service to initiate the connection to the remote host.
The socket layer will then request the connection from the transport layer. Once the connection is established, the application process will be unblocked with the correctly configured socket, and the guest application can proceed unaware of the behind-the-scenes process.
To support unmodified datacenter applications, \Boxer socket layer must implement a mechanism to support the complete (stream) socket interface with correct error handling, including non-blocking I/O, and interactions with other system features, such as the ability to share sockets between different processes. %

Figure~\ref{fig:listen-socket}, shows a subset of internal data structures on the passive side of the socket layer. 
It shows the state configured for 2 listening sockets. 
One of the sockets is shared between two guest application processes (P1 and P2), which are both waiting in blocking \texttt{accept} to receive new connected sockets.
Process P3 has a different socket and uses non-blocking I/O to accept new connections. 
However, both of the listening sockets are bound to the same local address.
This example is not an uncommon interaction of features used by datacenter applications, and \Boxer must be able to handle it.

The socket layer keeps track of sockets used by the guest processes.
First, \Boxer maintains the mapping between inodes and sockets in the application-socket-table, which can be used to uniquely identify each socket in the system.
When a process monitor sends a service request to the network service, first, it may need to look up the inode associated with the relevant socket and use that in the request. 
The socket layer can then map it to the unique socket data structure. In Figure~\ref{fig:listen-socket}, both processes P1 and P2 request \texttt{accept} service on the same inode value \texttt{i1} which maps their requests to the same listening socket entry.
Because these two requests are blocking, the PMs will block waiting on the responses. If there are no new connections available, the socket layer adds the service connections to the accept-queue of the \texttt{app-s-1} socket record, keeping the processes blocked. The accept queue will be drained once there are matching new connections that can be passed back to the blocked PMs, which will then return the new sockets to the guest processes.
Each listening socket record contains a reference to a connection-queue that will accumulate new matching connections, in the Figure~\ref{fig:listen-socket} example, all sockets point to the same connection-queue \texttt{connection-q-1}.
Connection-queues are created when guest processes create listening sockets bound to a new address. They are added to the connect-queue-table that is indexed by the listening address. In the example, there is only one connection-queue because both sockets \texttt{app-s-1} and \texttt{app-s-2} are listening on the same address \texttt{addr1}.

When the transport layer makes a connection request to the socket layer, it will use the connection destination address in the request.
This address is then used to lookup a matching connection-queueu, if one is found, it means that there is a guest process listening for such connection, and transport setup may continue. 
If there is no match, the request is denied, and the transport layer can propagate the error to the active side, potentially resulting in the (remote) client process receiving a connection refused error.

As new connections in a connection-queue become ready, references to to the matching listening sockets are used to return the new sockets to the blocked processes on the accept queues (e.g., Process 2 on the accept queue of socket app-s-1). The PMs are unblocked returning the new sockets to the application.

To handle non-blocking accept requests by the guest processes, when a new connection is available, the network service will create a new signal-connection to the local address that is bound to the real socket that the guest process is listening on.
This is also the socket that the guest process (oblivious to what is actually happening) will add to its I/O notifications (e.g., epoll\_ctl) to be notified by the kernel if there are new connections to accept.
The signal-connection is configured to trigger this event, and if the guest process chooses to accept, the PM will hide (and discard) the signal-connection and make a request to the network service to retrieve the new connection from the appropriate connection-queue, or return immediately if none are left.
\\

\noindent
\emph{Transport Layer}
This layer is responsible for the setup of data delivery for the sockets managed by the socket layer.
Currently, \Boxer has implementations of direct TCP, NAT-hole-punching TCP transport, and IP-forwarding-proxy TCP transport. 
Other transports, such as those based on S3, DynamoDB, or other intermediary services or overlays, could be implemented in the future.
The transport layer implementations use the control network managed by the NSs to exchange necessary messages to configure connectivity. For example, the NAT-hole-punching TCP transport that \Boxer uses in AWS Lambda, exchanges messages with remote \Boxer nodes to agree on the addresses to be used for NAT-hole-punching. Once agreed, direct TCP connections are established and passed up to the socket layer and then to the guest processes, transparently to the application.

\subsubsection*{Coordination Service}
\label{sec:Boxer-exec}
As \Boxer nodes join the network, they first contact a node that is the seed coordinator to be assigned a unique node ID, bootstrap their network membership set, and register their name.
All \Boxer nodes run a coordinator service that listens for membership updates to maintain its local membership set and to propagate updates to other nodes connected to it.
The membership set contains records for node-ids, their addressable IP addresses, and the (optionally) assigned names.

The NS can be configured to listen to the coordination service and only start executing its guest application when a certain number of nodes are present in the network or when a minimum number of nodes with specified names are present (e.g., only when a predefined number of workers are ready).
When the supervisor starts the guest application, it populates a set of local files with a list of other nodes, names, and node ids and the node id of the local node. Some guest applications can use these static files as part of their configuration. In addition to the static files, guest applications can use a local Unix domain socket interface to connect to the coordination service and stream dynamic membership updates.

\subsubsection*{Name Resolution}
\label{sec:Boxer-ns}
Names assigned to nodes in a \Boxer network are transparently available to the guest applications.
Guest processes that use standard system C Library name resolution functions that are intercepted by the program monitor, such as \texttt{getaddrinfo}, will transparently query the coordinator service for matches.
If the coordinator service produces no matches, the name resolution is forwarded to the underlying host and follows the standard path.
Other than the assigned names, the coordinator resolver also provides some canonical hostnames that can ease application configuration, e.g., `node-ID` name will always resolve to the IP address of the \Boxer node with the named ID.

\subsubsection*{Utilities}%
\label{sec:Boxer-fs}
\Boxer provides additional useful utilities, one of which is the ability to transparently remap file system names visible to guest applications.
This is useful when applications expect hard-coded pathnames that are not available or are restricted in FaaS.
\Boxer also uses this mechanism to redirect the application's accesses to some  `/etc/` configuration files that are read-only in FaaS environment (e.g.,\Boxer replaces `/etc/resolv.conf` with custom resolver configurations.)

\subsection{Container Orchestration with \Boxer}
\label{sec:boxer:deployment}

\begin{figure}[t]
  \centering
  \includegraphics[width=0.8\linewidth]{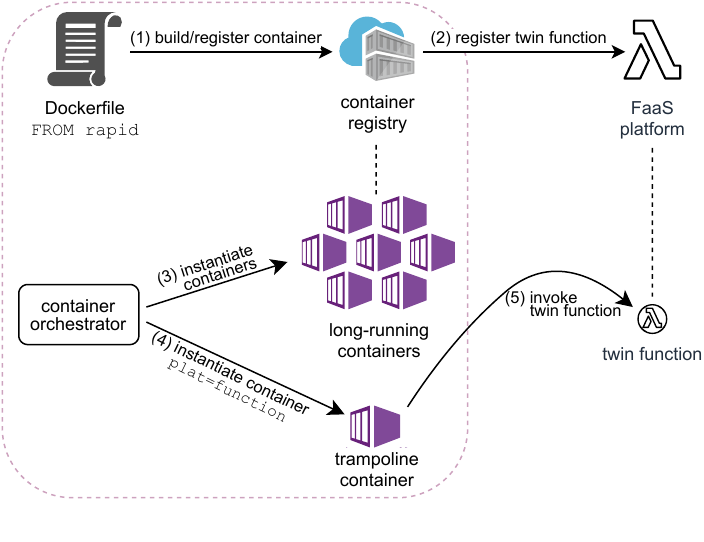}
  \caption{Container Orchestration with \Boxer.}
  \label{fig:exec-env}
\end{figure}

Container-based orchestration systems, such as Kubernetes~\cite{kubernetes}, Docker Swarm~\cite{swarm}, or Docker Compose~\cite{docker-compose}, are a common way to deploy and manage conventional datacenter applications. 
Therefore integrating \Boxer with these existing orchestration frameworks improves usability, lowers the barrier to adoption, and reduces the level of \Boxer-specific customization needed.
To enable this we produce \Boxer versions of commonly used base container images that can be used to transparently define application containers as if \Boxer was not present, and use \emph{trampoline container} technique to invoke \Boxer functions or containers via the same container orchestration systems (Figure~\ref{fig:exec-env}).
\Boxer trampoline containers are context-sensitive containers that start the container execution differently depending on their environment. 
When the orchestrator starts a container but specifies target platform to be a function, the container entrypoint does not start the \Boxer application in the container; instead, it collects the environment variables and the specified run command and invokes the corresponding twin function, passing the serialized environment as the invocation event, which is then used by function NS to join the overlay and start the appropriate container entrypoint running in FaaS.
This results in a new \Boxer node being added and the application running in the function, not in the original container.
The original container remains running as a phantom-container, receiving logs, waiting for the function container to terminate, signaling termination to the container orchestrator that stays under the illusion that the application is running locally.

\section{Evaluation}\label{sec:microbenchmarks}

\begin{figure}[t]
  \centering
   \begin{minipage}[t]{.235\textwidth}
    \includegraphics[keepaspectratio,width=1.08\textwidth]{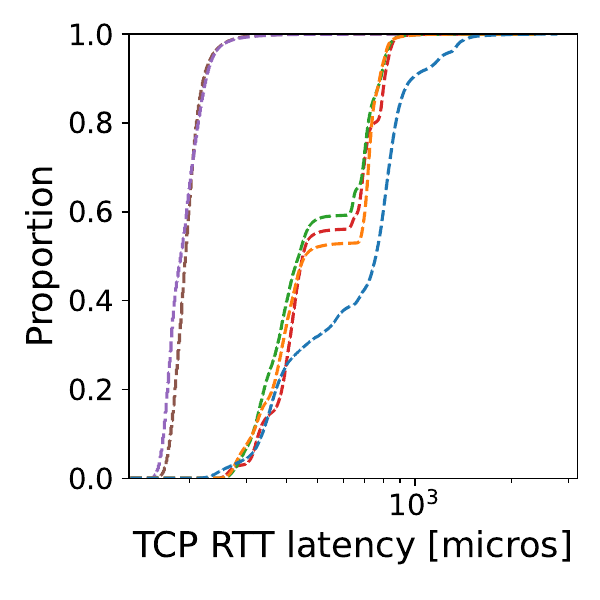}
    \centering
  \end{minipage}
  \begin{minipage}[t]{.235\textwidth}
    \includegraphics[keepaspectratio,width=1.08\textwidth]{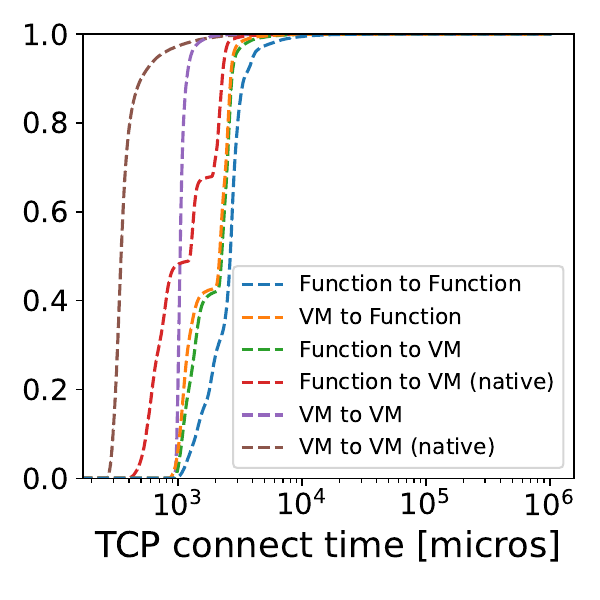}
    \centering
  \end{minipage}                     
   \caption{Empirical CDF of
  RTT latencies (left) and
  TTFB TCP connection establishment times (right)
   between different types of connecting and accepting hosts running \Boxer or without (native).}
  \label{fig:eval_tcp_connect}
  \label{fig:eval_tcp_latency}
    \label{fig:eval_network}
\end{figure}

\label{eval:deathstar}
To evaluate 
\Boxer 
we run, unmodified, DeathStarBench \cite{DeathStarBench}, a suite of cloud microservice benchmarks deployed using container networks that mimic how large scale, complex distributed applications are often deployed in the cloud.
We use the Zookeeper benchmark to demonstrate how \Boxer helps to quickly recover from node failures.

\subsection{Microbenchmarks}
To confirm that the properties of \Boxer provided networking are compatible with the ephemeral elasticity use case, we measured connection establishment times and latency of \Boxer provided TCP-hole-punching network transport for different combinations of endpoints.
We used EC2 m4.large VMs and Lambda Functions with 3007MB of memory as endpoints.
To measure connection establishment times, we measured time-to-first-byte (TTFB) observed by the client guest process running in \Boxer connecting to a remote server guest process also running in \Boxer. For comparison, we measured the native times without using \Boxer for the two possible combinations of functions connecting to VM and VM-to-VM directly.
To measure latency, the client program measured 128 rounds of 1024 byte ping-pong exchanges on an established connection.
For each endpoint combination, the experiments were repeated 1024 times on 32 distinct endpoint pairs.
Figure~\ref{fig:eval_network} shows the empirical CDF of the measurements.
As expected, the connection establishment times when using \Boxer TCP-hole-punching network transport compared to native times show the overhead of the connection setup and extra message round. 
Mean TTFB of VM-to-VM connections without \Boxer (native) is $408\mu s$ while with \Boxer is $1067\mu s$.
The latency results confirm that \Boxer adds no data path overhead once connections are established. The mean RTT for VM-to-VM connections with and without \Boxer are very similar at $198\mu s$ and $194\mu s$, respectively, and have similar distributions.
\Boxer provided Function-to-Function connections have mean TTFB connection establishment of $2735\mu s$, and RTT latency of $694\mu s$ with an increased dispersion. We find these acceptable for our use cases, especially considering that without \Boxer the application processes cannot establish Function-to-Function connections at all.

\subsection{Running DeathStarBench on \Boxer}
\label{sec:eval:deathstar}
Next, we focus on DeathStarBench's \textit{socialNetwork}, which offers a social network service to users and is organized using three microservice layers: i) front-end layer (implemented using an NGINX webserver); ii) logic layer (implemented using stateless Thrift services that communicate through RPCs); iii) caching and storage layer (implemented with MongoDB and Memcached instances). In  \textit{socialNetwork}, user requests are received by the front-end layer (NGINX web server) and then routed to one of the services in the logic layer. Depending on the user request, the logic layer may perform one or multiple requests to the caching and storage layers. Since the logic layer is stateless (i.e., it contains no internal persistent state), it can be deployed on AWS Lambda.

We did not have to make \emph{any modifications} to the application code to deploy DeathStarBench on AWS Lambda with \Boxer. The benchmark was only modified to i) use hostnames instead of fixed local IPs (for example, replace \texttt{192.168.1.7} by \texttt{nginx-thrift}), and ii) run all of the components of the front-end and logic layers using \Boxer. %

\begin{figure}[t]
  \centering
    \includegraphics[keepaspectratio,width=0.95\linewidth]{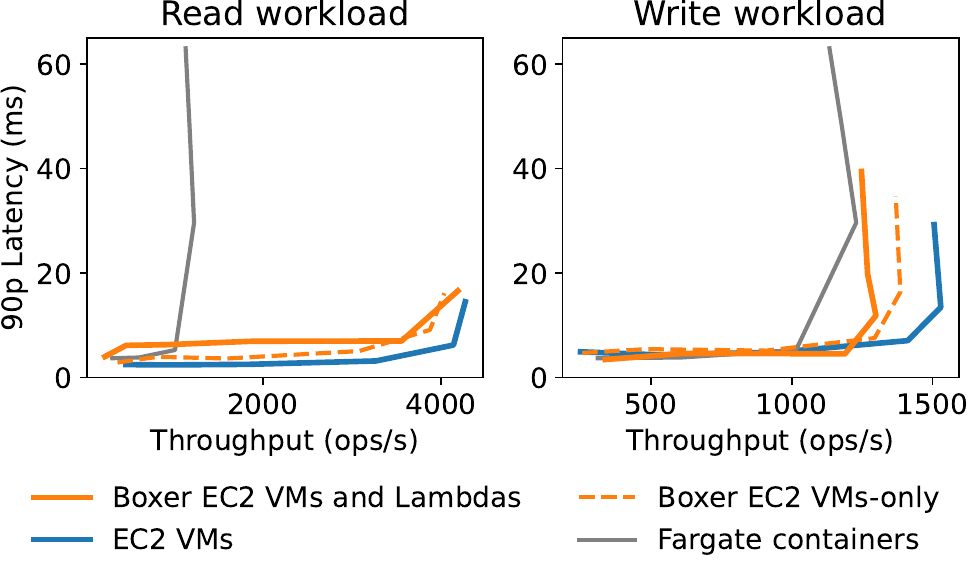}
   \caption{DeathStarBench results in a static deployment.}
  \label{fig:eval_deathstar_overhead}
    \label{fig:eval_deathstar}
\end{figure}

\subsubsection*{Methodology}
To evaluate %
\Boxer, we use four deployments (refered to as \textit{EC2-VMs}, \textit{\Boxer-EC2-VMs-only}, \textit{\Boxer-EC2-VMs-and-Lambdas}, \textit{Fargate-containers}), one baseline and three exercising \Boxer in different ways. (1) All components deployed as EC2 VMs(baseline, \textit{EC2-VMs}). (2) All components deployed as VMs in EC2 but the components the front-end and logic layers use \Boxer (\textit{\Boxer-EC2-VMs-only}). This deployment measures the performance overhead of using \Boxer. (3) A mixed deployment with front-end, and caching and storage layers are deployed as VMs, and logic layer deployed using Lambdas (\textit{\Boxer-EC2-VMs-and-Lambdas}). (4) A mixed deployment with the logic layer using AWS Fargate container service (\textit{Fargate-containers}).

To measure the throughput and latency of the end-to-end system we use two workloads included in the DeathStarBench suite. A read workload that issues requests to read a user timeline in the socialNetwork, and a write workload that creates follow relationships between users. Both workloads are generated using the \texttt{wrk} \cite{wrk} tool which builds and issues requests to the front-end layer. The performance of both workloads (read and write) is reported separately as each workload stresses the \Boxer overlay in a different way. The read workload mostly transfers data from the caching and storage layer (VMs), to the logic layer (VMs or Lambdas), and then to the front-end layer (VMs). The write workload operates in the opposite direction.

All experiments in this section were conducted in AWS Ohio (us-east-2) region. All VMs use a base Amazon Linux 2 \cite{amazon-linux}. For front-end, and caching and storage layers, we use \texttt{t3a.micro} instances due to the memory requirements of the services included in these layers. For the logic layer, when deployed in VMs, we use \texttt{t3a.nano} instances. 
Each Lambda is configured with 2048MB of memory (we experimentally determined that in us-east-2, the performance of a 2048MB Lambda is similar to a t3a.nano VM instance). For Fargate, we deploy containers also with 2048MB of memory and 1.0 vCPU unit (this configuration is also the one that yields faster container startup time, see Figure~\ref{fig:motivation_vm_startup}).

\subsubsection*{Overhead of using \Boxer}
Figure \ref{fig:eval_deathstar} shows the results for both read and write workloads across the four different types of deployments. For each workload, we collect the average throughput and 90th percentile latency with an increasing load in the system.
\Boxer introduces only a small overhead. For the read workload, the EC2 deployment becomes saturated at 3270 ops/s while the \Boxer-EC2-only becomes saturated at 3070 ops/s. For the same data points, the 90p latency of a single request for the EC2 and \Boxer-EC2-only deployments are 3.18 ms and 5.07 ms, respectively. Note that these latencies are measured end-to-end, thus include multiple internal microservice to microservice requests. The write workload demonstrates similar results. The EC2 and \Boxer-EC2-only deployments become saturated at 1411 ops/s and 1294 ops/s, with latencies of 7.07 ms and 7.56 ms, respectively.

We use a similar analysis to measure the overhead of launching the logic layer services in AWS Lambda by comparing the \Boxer-EC2-only and \Boxer deployments. Figure~\ref{fig:eval_deathstar_overhead} shows that for the read workload, the \Boxer deployment saturates at 3556 ops/s with a 90p latency of 7 ms. For the write workload, the same deployment saturates at 1189 ops/s and with a 90p latency of 4.55 ms.

We conclude that using \Boxer incurs a small performance overhead.
Moving services to Lambda also incurs a small overhead due to the different ways CPU and network are allocated to VMs and Lambdas. One could increase the memory budget assigned to lambdas to increase their vCPU allocation and thus close the gap between \Boxer-EC2-only and \Boxer.

\begin{figure}[t]
  \centering
    \includegraphics[keepaspectratio,width=1.0\linewidth]{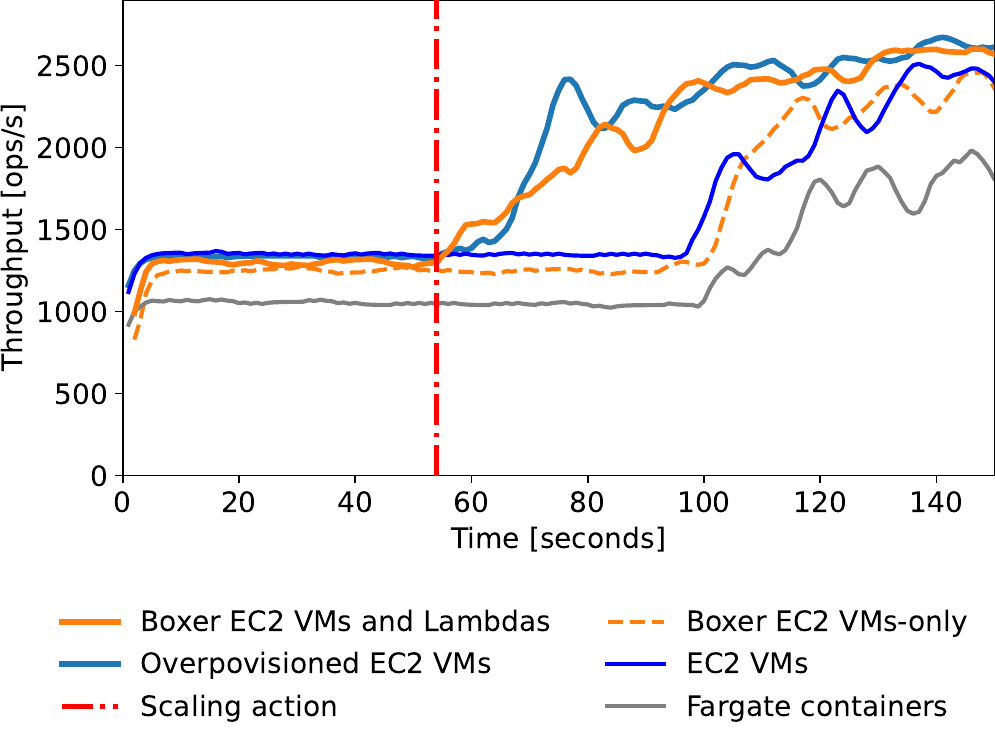}
  \caption{DeathStarBench write workload comparing elastic deployments (Section~\ref{eval:dynamic}).}
  \label{fig:eval_deathstar_dynamic}
\end{figure}

\begin{figure}[t]
  \centering
    \includegraphics[keepaspectratio,width=1.0\linewidth]{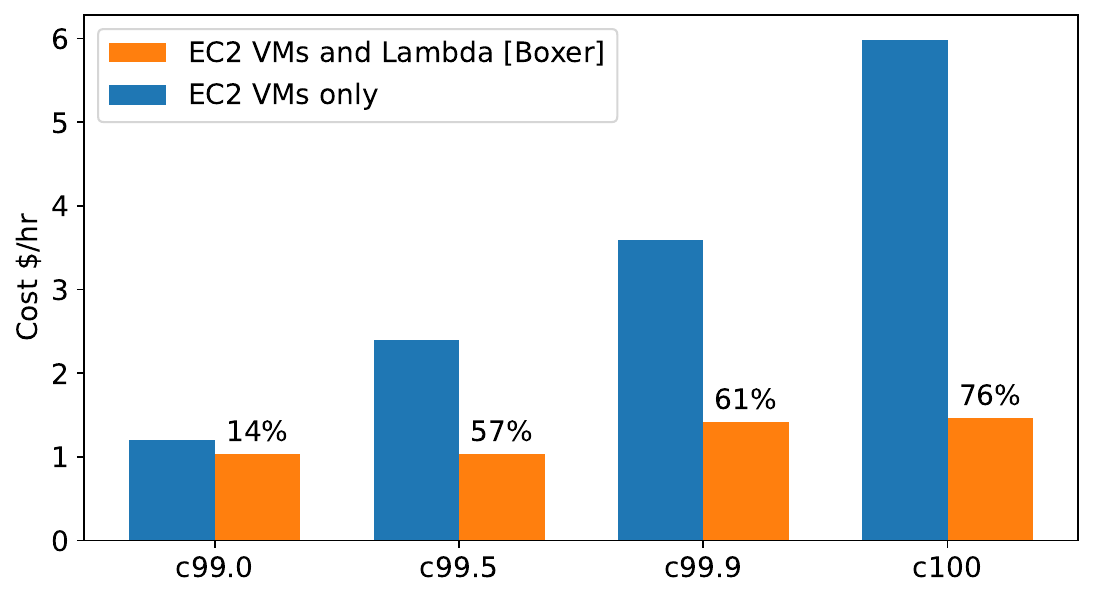}
  \caption{DeathStarBench logic layer absolute cost and cost reduction based on 1-day Reddit trace sample (Section~\ref{eval:dynamic}).
  }
  \label{fig:eval_deathstar_dynamic_cost}
\end{figure}

\subsubsection*{Elasticity through \Boxer}\label{eval:dynamic}

We now show how \Boxer can provide \emph{ephemeral elasticity} to increase the elasticity of microservices running on VMs and containers by leveraging serverless platforms. We start by deploying all logic layer services on VMs. When the load increases, additional logic layer services are allocated to handle the increased load either on VMs, containers, or Lambdas. In addition, we also include an overprovisioned VM deployment (\textit{Overp. EC2}) in which already allocated resources are added to the pool of workers. Our goal is to compare the elasticity of different deployment types.

Figure~\ref{fig:eval_deathstar_dynamic} presents a throughput trace for different deployments. Throughput is measured using wrk \cite{wrk} by looking at how many requests the front-end layer can handle per second. The tool dynamically increases the throughput based on the perceived system capacity. After approximately 55 seconds (dashed vertical line), a scaling action is taken to add a total of 12 workers to the pool of workers in the logic layer (one extra replica for each service in the logic layer). EC2 and Fargate take approximately 45 seconds to fully deploy all new workers (t=100s), while Lambda and overprovisioned EC2 scale almost immediately (approximately 1 second, t=55s). Using \Boxer to accommodate bursts reduces the time to add new workers to the pool by approximately 45$\times$ compared to EC2 and Fargate, providing comparable performance to VM-based overprovisioning.

Figure~\ref{fig:eval_deathstar_dynamic_cost} shows comparison of cost for using EC2 VM-based overprovisioning and \Boxer elasticity (using EC2 and Lambda). Based on a 1-day Reddit trace sample and the DeathStarBench throughput benchmarks (Figure~\ref{fig:eval_deathstar}), we calculated the necessary VMs for the logic layer to be overprovisioned to handle at least 99.0, 99.5, 99.9, and 100 percentile of requests/s in the trace (EC2-only). We then compared it to the cost of allocating a single VM instance for each logic layer service in VMs and on-demand dynamically scaling up using \Boxer to Lambdas to absorb load bursts. We observe that the cost reduction for using \Boxer provided elasticity ranges from 14\% to 76\% depending on the capacity levels.

\subsection{Elastic Fault Tolerance in Zookeeper}
\label{eval:zookeeper}
\begin{figure}[t]
  \centering
  \includegraphics[width=1.0\linewidth]{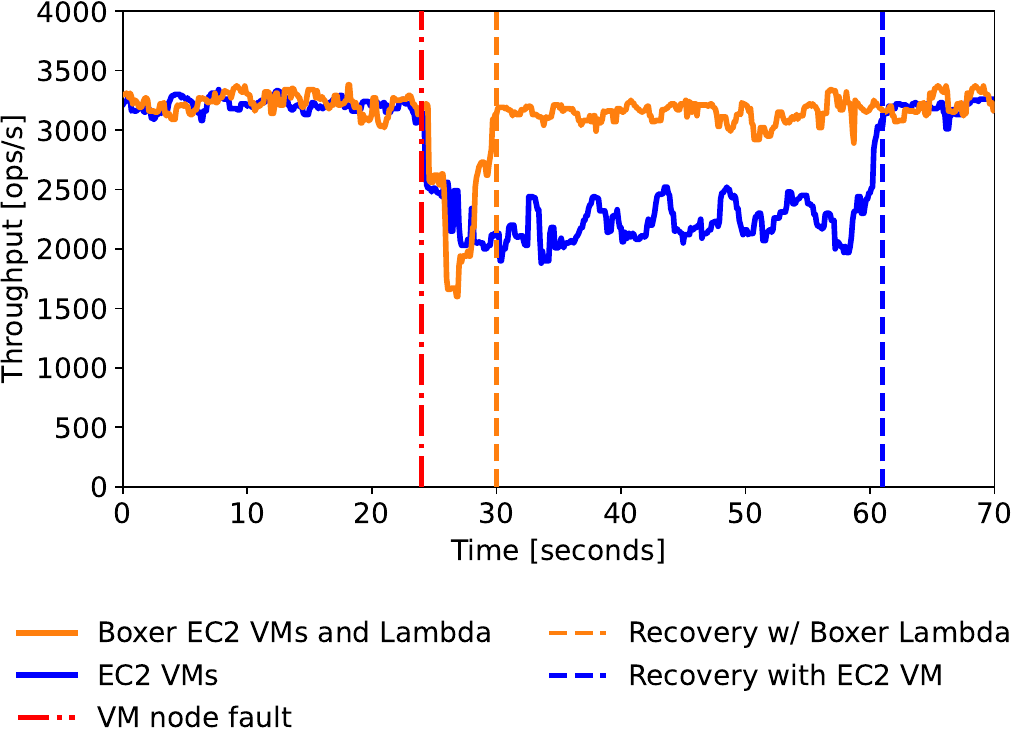}
  \caption{
  Recovering from node crash in a 3-node EC2 Zookeeper cluster using EC2 and  Lambda using \Boxer.
}
  \label{fig:eval_zookeeper}
\end{figure}
\Boxer provided elasticity can also be used to reduce down time due to recovery from node crashes. Minimizing node down time is crucial in highly dependable systems such as Zookeeper as read throughput drops and system guarantees may become compromised if additional faults happen before the first one is recovered. Moreover, having larger Zookeeper clusters to accommodate more faults is not common as write throughput degrades with additional replicas. %

For this scenario, we setup a 3-node Zookeeper~\cite{zookeeper} cluster deployed on EC2. Using this cluster, we forcibly shutdown one of the nodes and recover from the fault using either a newly allocated EC2 VM or using a Lambda with \textit{\Boxer}. We use \texttt{t3a.micro} VMs as Zookeeper nodes and Lambdas with 2048 MBs. We configure Zookeeper to allow dynamic reconfiguration, i.e., automatically adapt the quorum every time a new node leaves or joins the network. \Boxer is used to transparently allow a Zookeeper node deployed in a Lambda instance to join the quorum. In this recovery scenario, the rapidly deployed Zookeeper node in the short-lived Lambda function stays active only while a more permanent replica is being instantiated. We use a read-only workload based on the Zookeeper Benchmark\footnote{https://github.com/brownsys/zookeeper-benchmark}.
Figure~\ref{fig:eval_zookeeper} shows an execution trace of the workload throughput through time. After approximately 25 seconds, one of the Zookeeper VMs is shutdown and a new instance (EC2 or Lambda) is started to replace the failed node.
Using \Boxer, the fault recovered in under 6.5 seconds compared to 37.0 seconds with VMs (EC2), a 5.7x improvement in Zookeeper node recovery time.

\section{Discussion}

\textbf{Opportunities.}
\label{discuss:opportunities}
The elasticity bottleneck shifts from resource allocation to the application. 
By leveraging the \\ %
ephemeral elasticity, datacenter applications can quickly gain access to new resources; however, that does not necessarily mean that the applications can leverage the resources as quickly as they become available.
For example, load balancers, or controllers, may be configured to rebalance their workload among workers at an interval that is too high.

\noindent
\textbf{Current limitations.}
\label{discuss:limitations}
Several limitations in the current prototype will be  addressed as part of future work:
\emph{(1)} 
The current implementation of the process monitor relies on the lightweight dynamic linking mechanism for interposition, which cannot handle applications issuing system calls directly. 
We are investigating techniques based on lightweight selective binary rewriting~\cite{bin-rewrite} to extend the generality of the process monitor to arbitrary processes. 
\emph{(2)} \Boxer has only been tested on AWS, we plan to support other cloud platforms, but there may be provider-specific logic to adapt to each platform.
\emph{(3)} The interfaces that \Boxer emulates have complex semantics and, combined with the additional constraints, make handling all of the cases challenging. We have successfully tested \Boxer with several complex systems, but we are aware of corner cases that we have not handled yet, and we are working to make the system more complete.

\section{Related Work}

Although initially designed for event-triggered stateless functions, FaaS is now being presented as the next generation of cloud computing~\cite{Berkeley-CACM}. Fouladi et al. demonstrated using FaaS as a supercomputer on-demand to run highly parallel jobs like video encoding and software compilation~\cite{gg, ExCamera}. The \texttt{gg} framework enables users to run applications on FaaS by providing an intermediate representation and SDKs with which users can express their application as a composition of lightweight, functional containers~\cite{gg}. %
\Boxer pursues a similar vision but rather than just accelerating trivially parallel jobs, it aims to leverage FaaS for elasticity acceleration of off-the-shelf cloud applications. \Boxer also helps generic applications adapt to node failures and load spikes without resorting to overprovisioning, similar to how MArk~\cite{mark-atc19} spills to FaaS to accommodate ML inference load spikes, Beehive~\cite{beehive} utilizes FaaS nodes to offload load spikes for JVM-based applications, and Pixels-Turbo~\cite{pixels-turbo} accelerates query processing of unpredictable workload spikes with FaaS.

Improving elasticity by making compute and memory resources more fungible is an active area of research~\cite{nu,quicksand,harvestVM}. However, we argue that cloud users can benefit from high elasticity and greatly reduce overprovisioning for their applications without waiting for cloud providers to evolve and optimize their underlying infrastructure.

Complementary to \Boxer, others have explored enabling general computation on FaaS by providing GPU support~\cite{nuclio, serverless-gpu, serverless-gpu-sharing},  familiar concurrency APIs~\cite{kappa}, transactional workflows~\cite{beldi, olive},  atomicity guarantees over shared storage~\cite{aft-shim}, and handling timeouts by checkpointing and generating continuation functions~\cite{kappa}. Other optimizations such as locality-oriented scheduling~\cite{locality-serverless}, cold-start reduction~\cite{sock}, and memory footprint optimizations~\cite{medes} are orthogonal to \Boxer as our system is implemented on top of such serverless architectures and benefits from such optimizations.
 
Prior work has addressed function networking limitations by using intermediaries %
to relay messages between functions. \textit{mu}~\cite{ExCamera} proposed a framework for parallel computation and communication across buffers and relaying messages between functions. 
Others ~\cite{pocket, Anna18, Locus, Lambada, starling} leveraged external storage to exchange data between functions.
Projects such as InfiniCache~\cite{InfiniCache} and \texttt{gg}~\cite{gg} use a proxy-based %
approach. Nat-hole-punching in AWS Lambda has been previously leveraged for data analytics and for function communication primitives by~\cite{Boxer-CIDR21, fmi}, but not to transparently provide network-of-hosts model to datacenter applications.

\section{Conclusion}
We presented \Boxer, a system that transparently improves cloud application elasticity.
We demonstrated that it is possible to unbundle the \emph{event-triggered functions} programming model of FaaS from its underlying microVM resources and to provide the \emph{network-of-hosts} programming model on top of them.
We showed that this enables running cloud applications using publicly available FaaS infrastructure, which can be used to temporarily augment long-running unmodified cloud applications with fast ephemeral elasticity.
We showed that the availability of such fast ephemeral elasticity provides elasticity-fill that can significantly reduce the level of overprovisioning required to react to dynamic load and failure recovery. %

\bibliographystyle{ACM-Reference-Format}
\bibliography{paper}
\end{document}